\pdfoutput=1
 \documentclass[aps,prl,reprint,twocolumn,showpacs,superscriptaddress,%
nobibnotes,nofootinbib]{revtex4-1}  

\usepackage{tabularx}
\makeatletter
\def\hlinewd#1{%
\noalign{\ifnum0=`}\fi\hrule \@height #1 %
\futurelet\reserved@a\@xhline}
\makeatother
 \usepackage{auncial}
\usepackage{scrextend}
\usepackage[T1]{fontenc}
\usepackage{relsize}
\usepackage{amsmath}
\usepackage{amssymb}
\usepackage{epsfig}
\usepackage{graphicx}
\usepackage{hyperref}
\usepackage{dcolumn}   
\usepackage{tabu}
\usepackage{boldline}
\usepackage{slashed}
\usepackage{multirow}
\usepackage{color}
\usepackage[normal]{subfigure}
\usepackage{rotating}
\usepackage[margin=0.9in,a4paper]{geometry}
\usepackage[table]{xcolor}
\usepackage{enumitem}
\usepackage[utf8]{inputenc}
\usepackage{colortbl}
\usepackage{array,multirow}
\definecolor{nicered}{rgb}{0.7,0.1,0.1}
\definecolor{nicegreen}{rgb}{0.1,0.5,0.1}
\definecolor{red}{rgb}{1.0, 0, 0}

\hypersetup{colorlinks,citecolor= nicegreen,linkcolor= nicered}

\def\eq#1{{Eq.~(\ref{#1})}}
\def\eqs#1#2{{Eqs.~(\ref{#1})--(\ref{#2})}}

\def\fig#1{{Fig.~\ref{#1}}}

\def\vev#1{\left\langle #1\right\rangle}
\def\abs#1{\left| #1\right|}

\def\Tr{\mbox{Tr}\,}

\def\gsim{\raise0.3ex\hbox{$\;>$\kern-0.75em\raise-1.1ex\hbox{$\sim\;$}}}
\def\lsim{\raise0.3ex\hbox{$\;<$\kern-0.75em\raise-1.1ex\hbox{$\sim\;$}}}

\def\mb[#1]{\mathbf{#1}}

\renewcommand{\bar}{\overline}

\definecolor{LightCyan}{rgb}{0.88,1,1}
\definecolor{piggypink}{rgb}{0.99, 0.87, 0.9}
\definecolor{applegreen}{rgb}{0.55, 0.71, 0.0}
\definecolor{darkpastelgreen}{rgb}{0.01, 0.75, 0.24}
\definecolor{green-yellow}{rgb}{0.68, 1.0, 0.18}

\newcommand{\beq}{\begin{equation}}
\newcommand{\eeq}{\end{equation}}
\newcommand{\beqa}{\begin{eqnarray}}
\newcommand{\eeqa}{\end{eqnarray}}

\def\DFSZI{{\sf DFSZ-I}}

\def\DFSZI-II{{\sf DFSZ-I,II}}

\newcommand{\Sec}[1]{ \medskip \noindent {\sl \bfseries #1}}

\begin{document}

\begin{flushleft}                          
\footnotesize IPPP/18/62, DESY 18-127, TUM-HEP-1152-18
\end{flushleft} 



\title{Axion mass prediction from minimal grand unification}

\author{Luca Di Luzio}
\email{luca.di-luzio@durham.ac.uk}
\affiliation{\normalsize \it 
Institute for Particle Physics Phenomenology, Department of Physics, Durham University, DH1 3LE, Durham, United Kingdom}
\author{Andreas Ringwald}
\email{andreas.ringwald@desy.de}
\affiliation{\normalsize \it 
Deutsches Elektronen-Synchrotron DESY, Notkestra\ss e 85, D-22607 Hamburg, Germany}
\author{Carlos Tamarit}
\email{carlos.tamarit@tum.de}
\affiliation{\normalsize \it 
Physik Department T70, Technische Universit\"{a}t M\"{u}nchen, James Franck Stra\ss e 1, 85748 Garching, Germany}

\begin{abstract}
\noindent
We propose a minimal realization of the Peccei Quinn mechanism in a realistic SU(5) model, 
where the axion mass is directly connected to the grand-unification scale. 
By taking into account constraints from proton decay, collider searches and gauge coupling unification,  
we predict the axion mass: $m_a \in [4.8, 6.6]$\,neV. The upper bound 
can be relaxed up to $m_a < 330$ neV, at the cost of tuning the flavour structure of the proton decay operators. 
The predicted mass window will be complementarily probed by the 
axion dark matter experiments ABRACADABRA and CASPER-Electric, 
which could provide an indirect evidence for the scale of grand unification 
before the observation of proton decay.

\end{abstract}
%

\maketitle

\vspace{-20mm}
\Sec{Introduction.} 
It is a widespread belief that the standard model (SM) of particle physics 
should break down at some intermediate energy between the electroweak and the Planck scale. 
The quantum numbers of the SM fermions, together with the apparent convergence of the SM 
gauge couplings at high energies, hint to a unified gauge dynamics around $10^{15}$ GeV. 
This scale is generically compatible with indirect constraints from the non-observation of proton decay, 
the smoking-gun signature of Grand Unified Theories (GUTs). 
The search for proton decay was vigorously pushed in the past decades, 
and has slowly reached its limits with the Super-Kamiokande (SK) observatory \cite{Miura:2016krn}. 
Planned large-volume facilities, such as Hyper-Kamiokande (HK) \cite{Abe:2014oxa}, 
will improve the bound on 
the proton lifetime by one order of magnitude in the next decade.  
Though fundamentally important, that translates only into a factor of two on the GUT scale. 

Another well-motivated framework which points to energies in between the electroweak and the Planck scale is associated with the 
Peccei-Quinn (PQ) solution of the strong CP problem \cite{Peccei:1977ur,Peccei:1977hh}, which predicts the axion as 
a low-energy remnant \cite{Weinberg:1977ma,Wilczek:1977pj}. The axion needs to be 
extremely light and decoupled, and in a certain mass range it is also a viable dark matter (DM) candidate \cite{Preskill:1982cy,Abbott:1982af,Dine:1982ah}.
The experimental program for axion searches is rapidly evolving, 
with many novel detection techniques and new experiments being proposed recently \cite{Irastorza:2018dyq}.    
It is reasonable to expect that a large portion of the parameter space predicted by the QCD axion 
will be probed in the next decade. 
From an experimental point of view, however, one of the main bottlenecks of 
axion DM searches (e.g.~those exploiting microwave cavities or nuclear magnetic resonance techniques) 
is the need to perform a fine scan in the axion mass 
in order to meet a resonance condition. Since the axion mass is not predicted by the PQ mechanism, 
any extra theoretical information which could pin-down precisely the axion mass 
would be extremely helpful for experiments. 

Following recent attempts to revive PQ-GUTs in SO(10) \cite{Ernst:2018bib} 
(see also \cite{Reiss:1981nd,Mohapatra:1982tc,Holman:1982tb,Bajc:2005zf,Bertolini:2012im,Altarelli:2013aqa,Babu:2015bna}),  
in this Letter we revisit the more minimal option of SU(5). 
The simplest implementation of the axion in non-supersymmetric\footnote{The reader might 
wonder why we care for the fine-tuning of $\abs{\theta_{\rm QCD}} \lesssim 10^{-10}$ 
and not for the electroweak-GUT hierarchy. A possible answer is that 
the strong CP problem is qualitatively different from the hierarchy 
problem,
and it is conceivable that the solution of the latter does not rely on a 
stabilizing symmetry (an interesting example is the possibility that 
a light Higgs might be selected by the 
cosmological evolution of the universe \cite{Dvali:2003br,Dvali:2004tma,Graham:2015cka}).}
SU(5) was proposed long ago by 
Wise, Georgi and Glashow (WGG) \cite{Wise:1981ry}. 
However, similarly to the original SU(5) model of Georgi and Glashow (GG) \cite{Georgi:1974sy}, the WGG model 
is ruled out in its minimal formulation because of gauge coupling unification and neutrino masses.  
An elegant and minimal way to fix both these issues in the GG model 
was put forth some years ago by Bajc and Senjanovi\'{c} \cite{Bajc:2006ia}, 
which add to the minimal GG field content a single Majorana fermion representation, $24_F$, 
transforming in the adjoint of SU(5).    
The extra degrees of freedom have the right quantum numbers 
to generate neutrino masses via a hybrid Type-I+III seesaw mechanism 
and ensure a proper unification pattern. 
In particular, the main observable emerging from detailed renormalization group 
analyses of the GG+$24_F$ model (see Refs.~\cite{Bajc:2006ia,Bajc:2007zf,DiLuzio:2013dda}) 
is a clean correlation between light electroweak triplet states (constrained by the Large Hadron Collider (LHC)) 
and the unification scale (constrained by SK).

Having in mind the possibility of narrowing the axion mass range within a minimal and realistic 
extension of the WGG model, we extend the latter with a $24_F$ in analogy to the 
GG+$24_F$ case. This is actually welcome also from the point of view of the GG+$24_F$ model, 
which lacks a DM candidate. 
Within the WGG model (or any realistic extension of it) 
the axion mass can be put in one-to-one correspondence 
with the proton decay rate, regardless of the fine details of gauge coupling unification.  
This allows us to extract a generic upper bound on the axion mass. 
Including also the detailed information 
from gauge coupling unification available in the WGG+$24_F$ model,
we are also able to set a lower bound on the axion mass 
from the non-observation of electroweak-triplet states at LHC, 
thus predicting the following axion mass window: $m_a \in [4.8, 6.6]$ neV, 
where the upper bound holds in the absence of tuning of fermionic mixing.  
Next, we provide the axion coupling to the SM fields and estimate the 
sensitivity of future axion DM experiments such as ABRACADABRA \cite{Kahn:2016aff}  
and CASPEr \cite{Budker:2013hfa,JacksonKimball:2017elr} in the relevant mass window.

\Sec{The WGG model.} Let us recall the main features of the WGG 
model \cite{Wise:1981ry}. While the fermion content is 
that of the original GG SU(5) \cite{Georgi:1974sy}, namely 
three copies of $\bar 5_F$ and $10_F$ comprising the chiral SM matter fields, 
the scalar sector is extended to include a \emph{complex} $24_H$ and \emph{two} fundamentals, 
$5_H$ and $5'_H$. The WGG Lagrangian can be written as  
$\mathcal{L}_{\rm WGG} = \mathcal{L}_{\rm kin} + \mathcal{L}_{Y} - V_H$, 
where $\mathcal{L}_{\rm kin}$ encodes the (gauge) kinetic terms, 
the Yukawa Lagrangian is schematically\footnote{Non-renormalizable operators or extra scalar representations 
are further required in order to correct the ratio between the masses of down quarks and charged leptons.}
\beq 
\label{Lyuk}
\mathcal{L}_{Y} = \bar 5_F 10_F 5'^*_H + 10_F 10_F 5_H + \text{h.c.} \, ,
\eeq
while the scalar potential (which we do not report here entirely) 
contains two non-trivial invariants which are affected by global re-phasings:  
\beq 
V_H \supset 5'^\dag_H 24^2_H 5_H +5'^\dag_H 5_H \Tr (24^2_H)  + \text{h.c.} \, .
\eeq
Note that the structure of the WGG Lagrangian resembles 
that of the DFSZ 
model \cite{Zhitnitsky:1980tq,Dine:1981rt}. In fact $\mathcal{L}_{\rm WGG}$ is invariant under the 
global U(1)$_{\rm PQ}$ transformation: 
$\bar 5_F \to e^{-i\alpha/2} \bar 5_F$, $10_F \to e^{-i\alpha/2} 10_F$, 
$5_H \to e^{i\alpha} 5_H$, $5'_H \to e^{-i\alpha} 5'_H$ and $24_H \to e^{-i\alpha} 24_H$.     

We have performed the minimization of the full scalar potential in \cite{Wise:1981ry} 
and computed in turn the particle spectrum. In particular, it can be shown that 
the vacuum expectation value (VEV) configuration 
\beq 
\vev{24_H} = V \tfrac{1}{\sqrt{30}} \, \text{diag} (2,2,2,-3,-3) \, ,
\eeq
breaks SU(5)$\times$U(1)$_{\rm PQ}$ down to the SM gauge group 
with a \emph{single} order parameter $V$.\footnote{The recent work \cite{Boucenna:2017fna}, 
which bears some analogies with our proposal, differs crucially in the fact that the PQ symmetry is broken by an 
SU(5) singlet and hence the axion mass cannot be predicted.}  
The axion, 
the (pseudo) Nambu-Goldstone boson of the global U(1)$_{\rm PQ}$, 
is dominantly contained in the phase along the SM singlet direction of $24_H$, i.e.
\beq 
24_H \supset \vev{24_H} \tfrac{1}{\sqrt{2}} e^{i a / V} \, .
\eeq 
A crucial point of the WGG model is that the mass of the heavy vector leptoquark
$V_\mu = (3,2,-5/6)$ mediating proton decay, 
\beq 
\label{mVWGG}
m_V = \sqrt{\tfrac{5}{6}} g_5 V \, ,
\eeq
(where $g_5$ denotes the SU(5) gauge coupling)
is directly connected to the axion decay constant\footnote{We neglect corrections depending on weak-scale VEVs. For a pedagogical introduction and 
practical recipes on how to compute axion properties in GUTs, see Ref.~\cite{Ernst:2018bib}.} 
\beq 
\label{faVN}
f_a = V / \hat{N} \, ,
\eeq
where $\hat{N}$ is the  
U(1)$_{\rm PQ}$-SU(3)$_C$-SU(3)$_C$ 
anomaly coefficient, e.g.~$\hat{N} = 6$ in the WGG model.

This implies a generic 
relation between the axion mass and the 
proton decay rate.
By means of chiral effective field theory techniques, we can recast the master formula for the proton decay mode 
$p \to \pi^0 e^+$ in SU(5) as \cite{Claudson:1981gh,Nath:2006ut}: 
\begin{align}
\label{Gammapdecay1}
\Gamma_{p \to \pi^0 e^+} &= \frac{m_p}{16 \pi f_\pi^2} A^2_L \abs{\alpha}^2 (1+D+F)^2  \nonumber \\ 
&\times  \left(\frac{g_5^2}{2 m^2_V}\right)^2 
\left[ 4 A^2_{SL}  + A^2_{SR} \right] \, ,
\end{align}
where we have set unknown fermion mixing rotations to a unit matrix 
(see \cite{Nath:2006ut} for complete expressions).
$A_L = 1.25$ encodes the renormalization from the electroweak scale to the proton mass, 
$m_p = 938.3$ MeV; $f_\pi = 139$ MeV, $D=0.81$, $F=0.44$ and $\alpha=-0.011$ GeV$^3$ are 
phenomenological parameters given by the chiral Lagrangian and the lattice. 
$A_{SL(R)}$ are short-distance renormalization factors from the GUT to the electroweak scale 
which depend on the intermediate-scale thresholds \cite{Buras:1977yy,Wilczek:1979hc}. 
Compact expressions for the latter 
can be found e.g.~in Ref.~\cite{Bertolini:2013vta}. For instance, running within the SM 
from $10^{15}$ GeV to the electroweak scale yields $A_{SL} = 2.4$ and $A_{SR} = 2.2$. 

By using \eqs{mVWGG}{faVN} and the relation 
$m_a = 5.7 \, \text{neV} \, (10^{15} \, \text{GeV} / f_a)$ \cite{diCortona:2015ldu,Borsanyi:2016ksw} 
we can re-express \eq{Gammapdecay1} in the following parametric form:
\begin{align}
\label{Gammapdecay2}
\Gamma_{p \to \pi^0 e^+} &\simeq \left( 1.6 \times 10^{34} \ \text{yr} \right)^{-1} \left( \frac{m_a}{3.7 \ \text{neV}} \right)^4 
\left( \frac{6}{\hat{N}} \right)^4
\nonumber \\ 
& \times \left[ 0.83 \left( \frac{A_{SL}}{2.4} \right)^2 + 0.17 \left( \frac{A_{SR}}{2.2} \right)^2 \right] 
\, , 
\end{align}
where we have highlighted in the first parenthesis 
the current proton decay bound from SK \cite{Miura:2016krn}.  
Remarkably, this translates into an upper bound for the axion mass which, although affected by the model-dependent parameter $\hat N$, is independent of the fine details of the unification analysis 
that enter only logarithmically into $A_{SL(R)}$. 

\Sec{\boldmath Axion mass prediction in WGG+$24_F$.} 
The failure of the WGG model in explaining neutrino masses and gauge coupling unification 
can be readily fixed by adding a single Majorana representation, $24_F$, in analogy to the proposal 
of Ref.~\cite{Bajc:2006ia}. 
Here, we highlight the main differences due to the presence 
of the PQ symmetry. The Yukawa Lagrangian is extended by 
\beq 
\label{DeltaLY}
\Delta \mathcal{L}_Y = \bar 5_F 24_F 5_H + 
\Tr 24^2_F 24^*_H  + \text{h.c.} \, .
\eeq
The first term provides a Dirac Yukawa interaction for the fermion triplet 
and singlet 
fields contained in 
$24_F$, while the second term generates a Majorana mass for the full multiplet 
upon SU(5) symmetry breaking. We leave implicit the presence of extra non-renormalizable 
operators which are needed for two reasons: $i)$ 
to avoid a rank-one light neutrino mass matrix and $ii)$ to split the mass of the $24_F$ sub-multiplets 
(for further details see \cite{Bajc:2006ia,Bajc:2007zf,DiLuzio:2013dda}). 
 \eq{DeltaLY} also fixes the PQ transformation of the new field: $24_F \to e^{-i\alpha/2} 24_F$;
including the latter the total 
U(1)$_{\rm PQ}$-SU(3)$_C$-SU(3)$_C$ anomaly 
yields $\hat{N} = 11$. 

The possibility of narrowing down the 
axion mass range follows directly from unification constraints. 
The main issue with gauge coupling unification in the SM is the early convergence of the electroweak 
gauge couplings, $\alpha_1$ and $\alpha_2$, around $10^{13}$ GeV, at odds with 
proton decay bounds. 
Hence, the key ingredients for a viable unification pattern are additional particles charged under 
SU(2)$_L$ which can delay the meeting of $\alpha_1$ and $\alpha_2$. 
Such a role in the WGG+$24_F$ model can be played by the electroweak fermion $T_F = (1,3,0)$ 
and scalar $T_H = (1,3,0)$ triplets contained in the $24_{F,H}$.\footnote{Compared to the GG+$24_F$ case 
we have in principle extra thresholds due to fact that the $24_H$ is complex. 
However, the constraints coming from the minimization of the scalar potential 
imply that only one \emph{real} triplet can be light, 
otherwise a colored octet scalar would be lowered to the triplet mass scale, spoiling nucleosynthesis  \cite{Bajc:2006ia}.} 
They are predicted to be at the TeV scale, 
so that a large enough unification scale can be achieved. 

Both types of triplets, if light enough, can give interesting signatures at the LHC. 
The fermionic component leads to same sign di-lepton events which violate lepton number \cite{Arhrib:2009mz}.   
A recent CMS analysis \cite{Sirunyan:2017qkz} 
sets a $95\%$ CL exclusion at $840$ GeV, while projected limits at the 
High Luminosity LHC (HL-LHC) \cite{Ruiz:2015zca,Cai:2017mow} 
give $m_{T_F} \gtrsim 2$ TeV. 
Bosonic triplets can affect the di-photon 
Higgs signal strength, but the bound is milder compared to the fermionic triplet and 
model-dependent \cite{Chabab:2018ert}. Here we assume a conservative $m_{T_H} \gtrsim 200$ GeV. 

The complete unification pattern including also the convergence of $\alpha_3$ 
with $\alpha_1$ and $\alpha_2$ requires heavier colored particles. These are the color-octet fermions 
and scalars contained in the $24_{F,H}$, whose masses are  required to be around $10^8$ GeV, 
well beyond the LHC energy range. 

The main prediction of gauge coupling unification is hence a clean 
correlation between a triplet mass parameter 
(whose analytical form is a consequence of the $\alpha_2$ beta function),
\beq 
\label{m3def}
m_3 = \left( m_{T_F}^4 m_{T_H} \right)^{1/5} \, , 
\eeq 
and the unification scale. The latter is operatively defined as the energy scale where $\alpha_1$ and $\alpha_2$ 
meet up to GUT-scale thresholds \cite{Weinberg:1980wa,Hall:1980kf}, 
and it can be identified with $m_V$, the mass of the heavy vector leptoquark $V_\mu$ 
mediating proton decay. Thanks to \eqs{mVWGG}{faVN},  
we can trade $m_V$ for the axion mass, 
which allows us to present the unification constraints in the $(m_a, m_3)$ plane. 

\begin{figure}
\includegraphics[width=.45\textwidth]{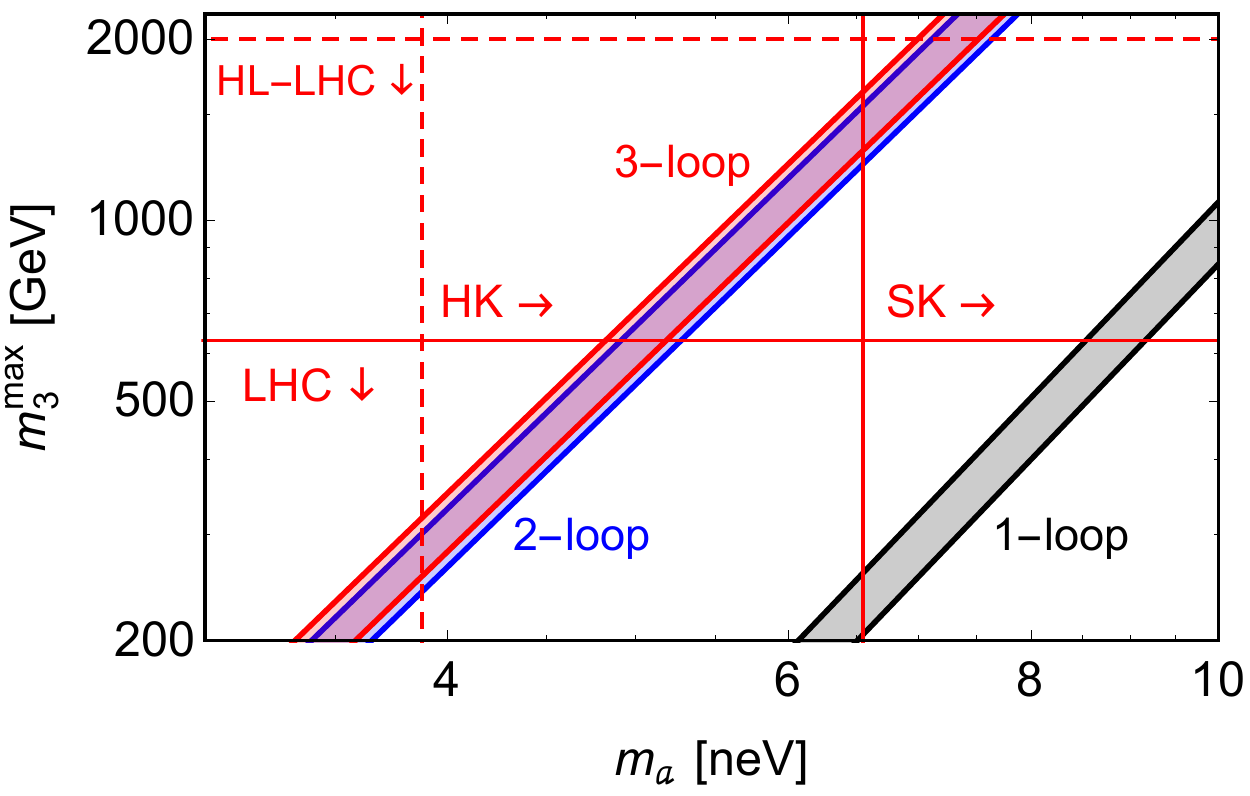}
\caption{\label{fig:m3vsma}
Maximal triplet mass parameter as a function of the axion mass. 
Grey, blue and red bands denote respectively the correlation at 1, 2 and 3 loops 
(shaded regions encode the 1$\sigma$ uncertainty on the electroweak gauge couplings). 
The full horizontal (vertical) red line is the current exclusion from LHC (SK), 
the dashed horizontal (vertical) red line is the projected exclusion from HL-LHC (HK). 
}
\end{figure}

Following Ref.~\cite{DiLuzio:2013dda}, we have performed 
a gauge coupling unification analysis including the leading NNLO corrections 
coming from the 2-loop matching coefficients and the 3-loop beta functions 
due to the fermion and scalar triplets. The extra thresholds affecting the evolution of 
$\alpha_1$ and $\alpha_2$ are fixed in such a way that the value of $m_3$ is maximized 
(cf.~\cite{DiLuzio:2013dda} for more details), which defines the parameter $m_3^{\text{max}}$.    
The results are displayed in \fig{fig:m3vsma} which shows the correlation in the $(m_a, m^{\text{max}}_3)$ 
plane. Taking into account the present bounds from LHC (on both fermion and scalar triplets) 
and SK (obtained by setting $A_{SL} = 2.6$ and $A_{SR} = 2.4$ in \eq{Gammapdecay2},  
which follow from the unification analysis), 
the preferred axion mass window is 
\beq 
\label{eq:preferred_window}
m_a \in [4.8, 6.6]\  \text{neV}\,.
\eeq
Future projections at HL-LHC 
(where we represent only the sensitivity to the fermion triplet mass)
and HK (10 years data taking \cite{Abe:2014oxa}) 
can complementary test this scenario. 

We remark that the SK bound was imposed 
via \eq{Gammapdecay2}, which
does not account for possible cancellations in the flavour structure of the proton decay operators. 
By considering different proton decay channels 
and accounting for flavour rotations,
one can still extract a model-independent bound on the unification scale 
which is about an order of magnitude smaller \cite{Dorsner:2004xa,Kolesova:2016ibq}. 
The absolute upper bound on the axion mass is obtained by tuning 
to zero all the main proton decay channels, except those involving strange mesons. 
Using the results of Ref.~\cite{Dorsner:2004xa} for the case of heavy Majorana neutrinos 
and updated with the latest
experimental limit $\tau / \mathcal{B} (p \to K^0 \mu^+) > 1.3 \times 10^{33}$ yr \cite{Regis:2012sn}, 
we obtain $m_a < 330$ neV. Similarly, from the projections at HK (10 years data taking \cite{Abe:2014oxa}) 
in the $p \to K^+ \bar \nu$ channel we estimate $m_a < 160$ neV.

\Sec{Sensitivity of future axion DM searches.} 
An axion in this mass range is extremely weakly coupled to SM particles, since its couplings 
to e.g.~photons ($\gamma$), electrons ($e$), protons ($p$), and neutrons ($n$) are 
inversely proportional to the axion decay constant,
\begin{equation}
{\cal L}_a \supset 
\frac{\alpha}{8\pi} \,\frac{C_{a\gamma}}{f_a}\,a\,F_{\mu\nu} {\tilde F}^{\mu\nu} 
-
\frac{1}{2}\, \frac{C_{af}}{f_a} \, 
\partial_\mu a\ \overline{\Psi}_f \gamma^\mu 
\gamma_5   \Psi_f \,.
\end{equation}
while the coefficients $C_{ax}$ are of order unity. 
In the WGG+$24_F$ model,  we find:
\begin{align}\label{eq:Cs}\begin{aligned}
C_{a\gamma} &= \tfrac{8}{3}-1.92(4) \,, \qquad
C_{ae} = \tfrac{2}{11}\sin^2\beta\,,  \\
C_{ap}&=-0.47(3) \\
&+\tfrac{6}{11}[0.288\cos^2\beta-0.146\sin^2\beta\pm 0.02]\,,  \\
C_{an}&=-0.02(3)  \\ 
&+\tfrac{6}{11}[0.278\sin^2\beta-0.135\cos^2\beta\pm 0.02]\, , 
\end{aligned}\end{align}
where we introduced the ratio of the electroweak VEVs,   
$\tan \beta = \vev{5_H} / \vev{5_{H'}}$. 
This makes the GUT axion clearly invisible for purely laboratory based experiments. 

However, axions in this mass range are known to be excellent DM candidates \cite{Preskill:1982cy,Abbott:1982af,Dine:1982ah}
which can be searched for in axion DM direct detection experiments. 
In fact, very light axion DM even tends to be overproduced and can only 
be reconciled with the measured amount of cold DM if the PQ symmetry remained broken during and after inflation in the early universe.\footnote{This solves at the same time  the cosmological SU(5) monopole problem and the PQ domain-wall problem 
(the WGG$+24_F$ model has domain-wall number 11).}
In this case, the relative contribution of axion 
DM to the energy density of the universe
depends not only on the mass, but also on the initial value of the axion field $a_i$ in units of the decay constant, $\theta_i = a_i/f_a$,  
inside the causally connected region which is inflated into our visible universe, cf.  
 \cite{Borsanyi:2016ksw,Ballesteros:2016xej}:
\begin{equation}
\Omega_ah^2 
= 0.12\,\left({ 5.0~{\rm neV}\over m_a}\right)^{1.165}\,
\left(\frac{\theta_{\rm i}}{1.6\times 10^{-2}}\right)^2  \,.
\end{equation}
Thus an axion in the neV mass range can make 100\,\% of DM, if the initial field value $\theta_i$ is of order 
$10^{-2}$.\footnote{This value can be supported by anthropic arguments  \cite{Tegmark:2005dy}.}  
In this cosmological scenario, however, 
quantum fluctuations of a massless axion field during inflation may lead to isocurvature density fluctuations that get imprinted in the temperature fluctuations of the cosmic microwave background (CMB) \cite{Linde:1985yf,Seckel:1985tj}, whose amplitude is stringently constrained by observations. In the case that the $24_H$ stays at a broken minimum of the potential throughout inflation 
(e.g.~for a SM-singlet inflaton), those constraints translate in an upper bound 
on the Hubble expansion rate during inflation \cite{Beltran:2006sq,Hertzberg:2008wr,Hamann:2009yf}: 
\begin{equation}
H_I < 5.7\times 10^8\,{\rm GeV} \left( \frac{5.0\,\text{neV}}{m_a}\right)^{0.4175}
\,.
\end{equation}
Intriguingly, these isocurvature constraints can disappear completely  in the case of non-minimal chaotic inflation 
\cite{Spokoiny:1984bd,Futamase:1987ua,Fakir:1990eg} along one of the components of the $24_H$. 
In this case, during inflation the $24_H$ is not at a minimum, Goldstone's theorem does not apply, and the lightest fluctuations orthogonal to the inflaton can have masses  above $H_I$ as long as the 
parameter $\xi_{24_H}$, describing the non-minimal coupling to 
the Ricci scalar, $S \supset - \int d^4 x\, \sqrt{-g}\, \xi_{24_H}\Tr (24^2_H) R$, is larger than $\sim 0.01$.  For $\xi_{24_H}$ above this value, the power spectra of the isocurvature fluctuations become exponentially suppressed and the CMB bounds can be avoided. In such scenarios, one still needs to ensure that the PQ symmetry is never restored after inflation; we expect that this might be possible for small enough quartic and Yukawa couplings of the 
$24_H$, but a dedicated analysis generalizing the non-pertubative and perturbative reheating calculations in Ref.~\cite{Ballesteros:2016xej} is needed.

\begin{figure}
\includegraphics[width=.45\textwidth]{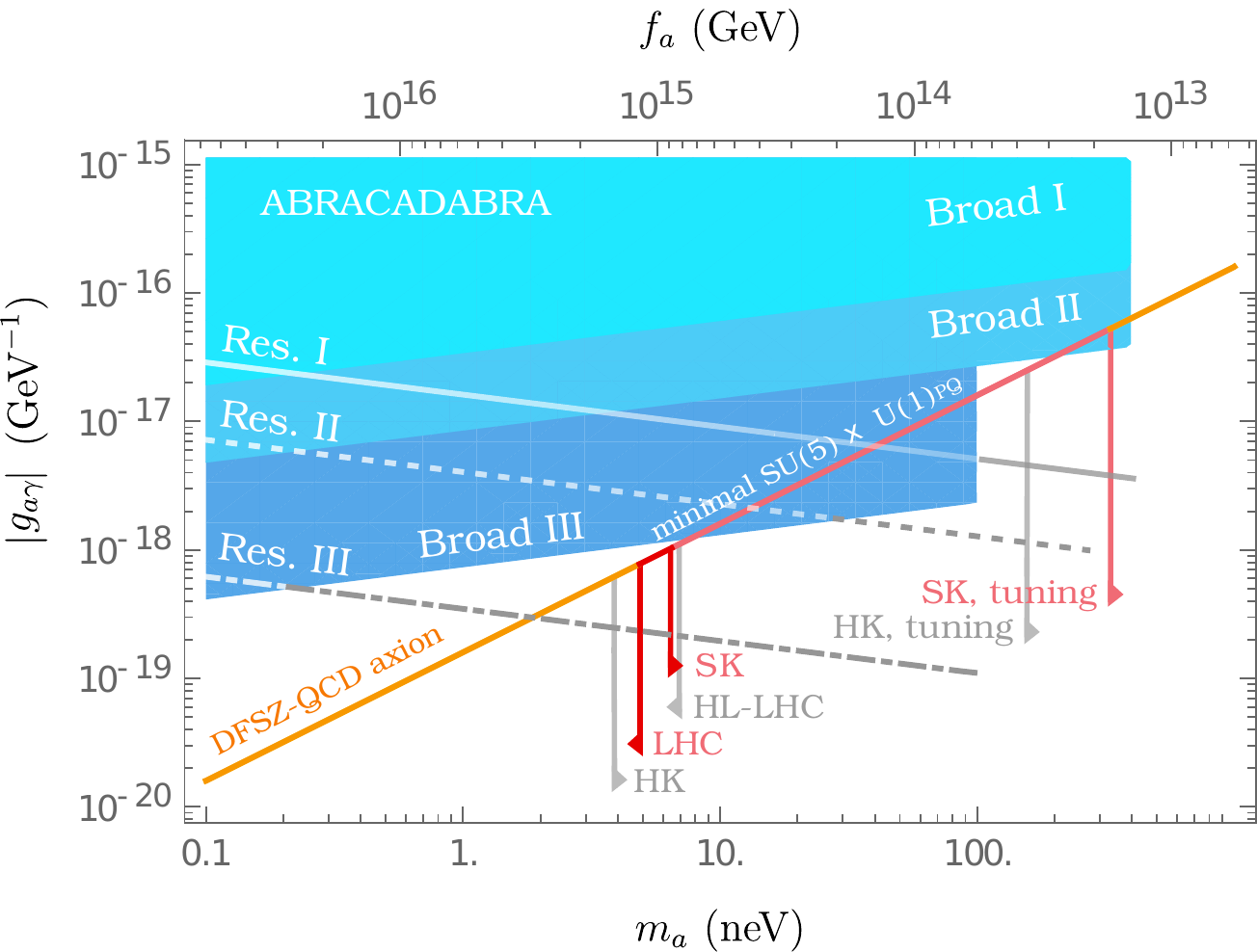}
\caption{\label{fig:gagg_vs_ma}
Axion coupling to photons, $g_{a\gamma}$, versus axion mass $m_a$. 
The blue regions give the projected sensitivities of  broadband (``Broad'') 
and resonant (``Res.'') search modes of ABRACADABRA
from Ref.~\cite{Kahn:2016aff}. 
}
\end{figure}
\begin{figure}
\includegraphics[width=.45\textwidth]{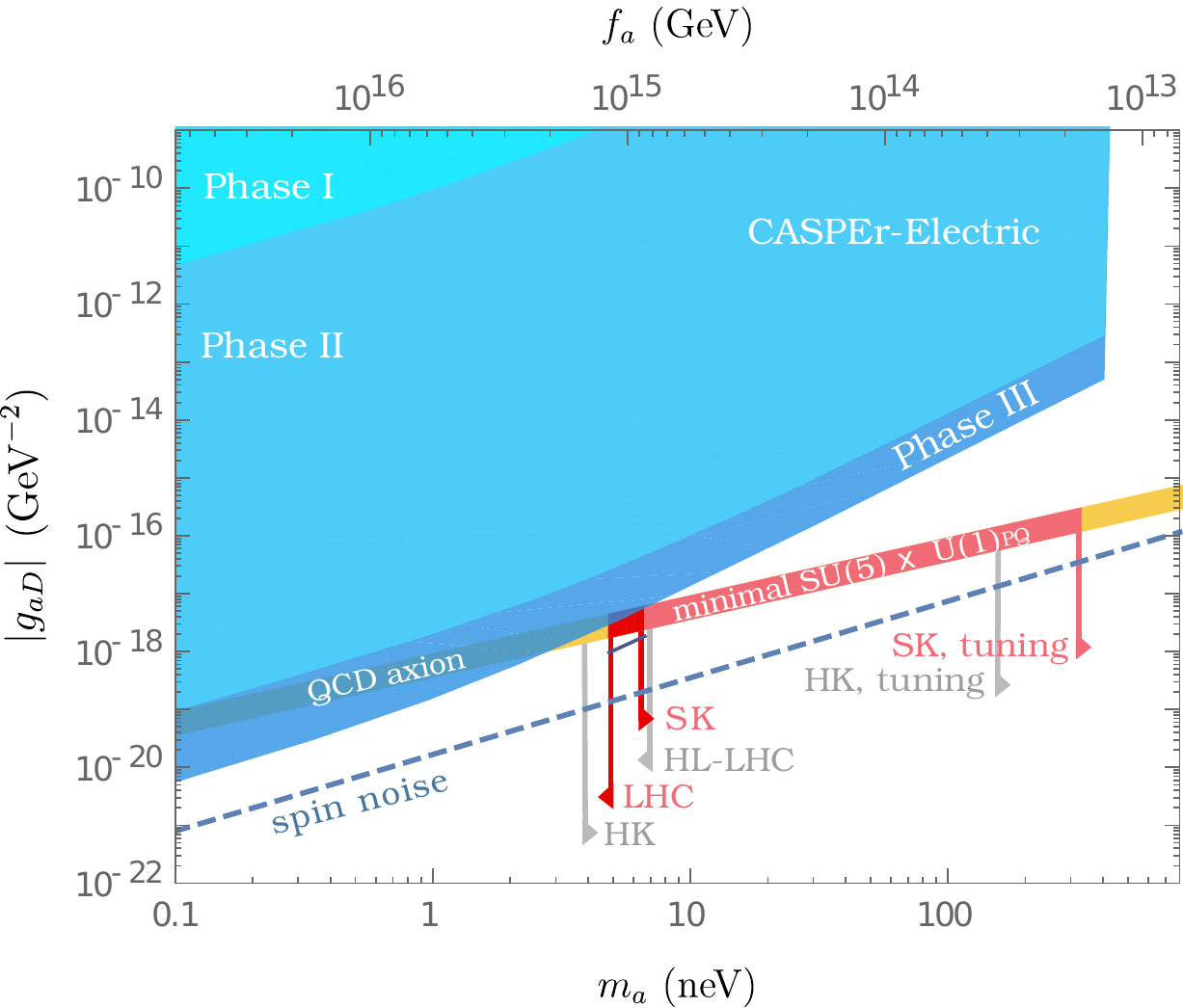}
\caption{\label{fig:gd_vs_ma3}
Axion coupling to the nucleon EDM operator, $g_{aD}$, versus axion mass $m_a$. 
The blue regions give the projected sensitivities of CASPEr-Electric 
from Ref.~\cite{JacksonKimball:2017elr}. The short, full blue line reflects a factor of three improvement 
in sensitivity for a search just concentrated on the preferred mass region.
}
\end{figure}
\begin{figure}
\includegraphics[width=.45\textwidth]{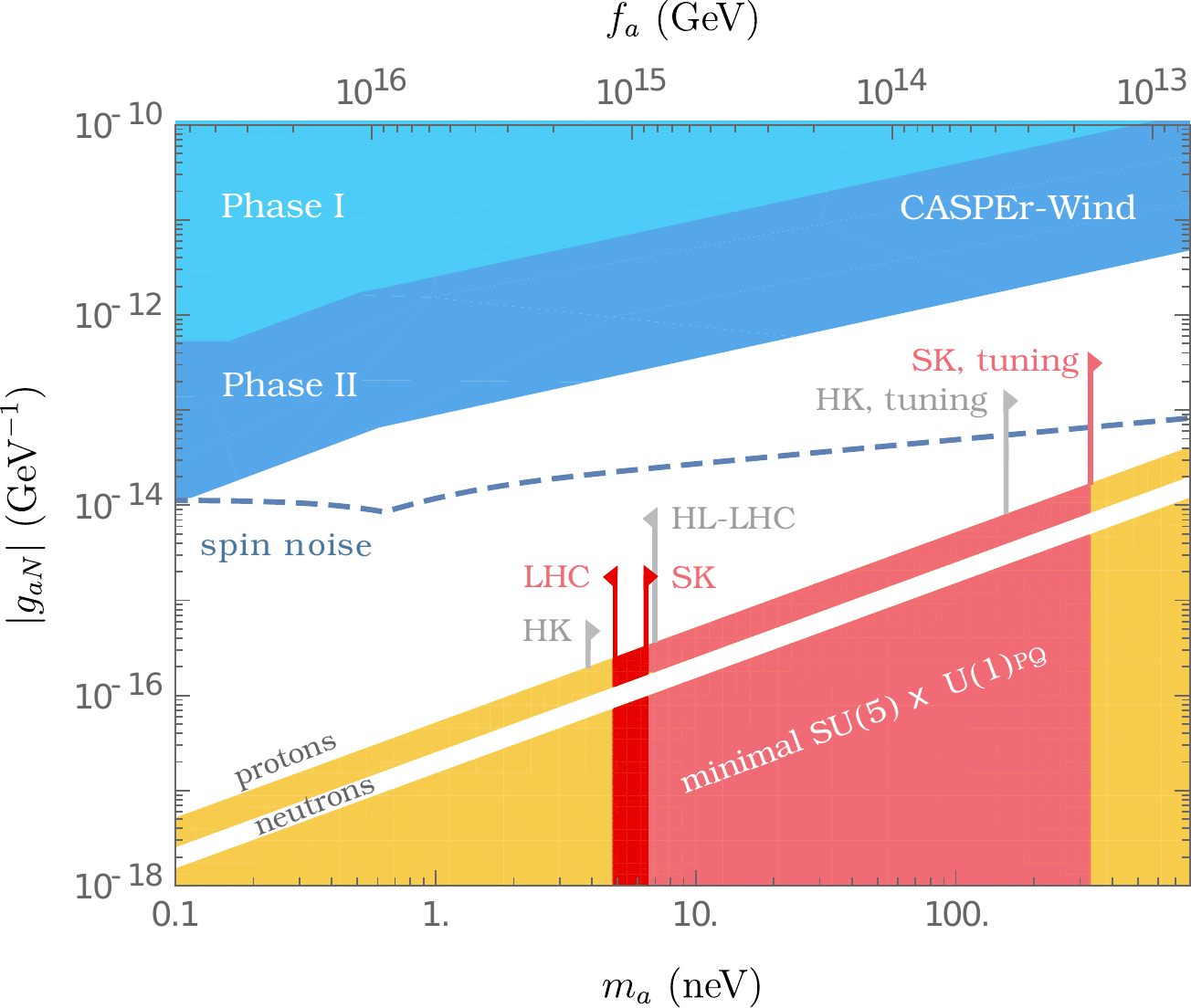}
\caption{\label{fig:gaNN_vs_ma}
Axion coupling to the nucleons, $g_{aN}$, versus axion mass $m_a$. 
The blue regions give the projected sensitivities of CASPEr-Wind 
from Ref.~\cite{JacksonKimball:2017elr}. 
}
\end{figure}

The DM experiment ABRACADABRA \cite{Kahn:2016aff}, has very good prospects to probe the 
axion photon coupling, $g_{a\gamma} = \alpha \, C_{a\gamma} / (2 \pi f_a)$, in the relevant mass region. 
This is shown in \fig{fig:gagg_vs_ma}, from which we infer that the whole parameter space 
of the WGG+$24_F$ model (including the tuned region) 
can be tested in the third phase of the broadband and resonant search modes of ABRACADABRA. 

In Fig.~\ref{fig:gd_vs_ma3}, we confront our axion mass prediction with the projected sensitivity 
of the experiment CASPEr-Electric \cite{Budker:2013hfa,JacksonKimball:2017elr}, which aims to search 
for oscillating nucleon electric dipole moments (EDM) $d_n (t) = g_{aD}\,\frac{\sqrt{2\rho_{\text{DM}}}}{m_a} \cos (m_a\,t)$
 \cite{Graham:2013gfa},
where $g_{aD}$ is the model-independent coupling of the axion to the nucleon EDM operator,
$\mathcal{L}_a \supset -\frac{i}{2} g_{aD}\, a\, \overline{\Psi}_N \sigma_{\mu\nu} \gamma_5 \Psi_N F^{\mu\nu}$,
and $\rho_{\rm DM}=0.3\,\text{GeV}/\text{cm}^3$ is the local energy density of axion DM. 
The QCD axion band in Fig.~\ref{fig:gd_vs_ma3} indicates the theoretical uncertainty of the non-perturbative estimates
of $g_{aD}$. We used the result in \cite{Pospelov:1999mv}, obtained with QCD sum rules; for other evaluations
see e.g.~\cite{Crewther:1979pi,Hisano:2012sc}.\footnote{Current lattice QCD results on $g_{aD}$ 
do not show a statistically significant non-zero signal \cite{Yoon:2017tag}.}
We infer from Fig.~\ref{fig:gd_vs_ma3}, that the preferred axion mass window \eqref{eq:preferred_window} could definitely 
be probed in phase III of CASPEr-Electric.\footnote{The sensitivity in $g_{aD}$ improves with the scanning time 
as $t^{1/4}$. This amounts to a factor of three improvement (denoted by a short, full blue line in \fig{fig:gd_vs_ma3}), 
if CASPEr-Electric spends all the measurement time 
just on the preferred mass region.} 

On the other hand, 
the projected sensitivity of CASPEr-Wind \cite{JacksonKimball:2017elr}, which  exploits the axion nucleon coupling 
$g_{aN}=C_{aN}/(2f_a)$ ($N=p,n$) to search for the axion DM wind due to the movement of the Earth through the 
Galactic DM halo \cite{Graham:2013gfa}, misses the preferred coupling vs.~mass region by two orders of magnitude or more, even in its phase II. We show this in \fig{fig:gaNN_vs_ma}, where the theoretical uncertainty of the axion band 
is obtained from the errors in the coefficients of \eq{eq:Cs}, and from varying $\tan\beta \in [0.28,140]$ 
in the perturbative unitarity domain \cite{Giannotti:2017hny}.

\Sec{Conclusions.}
In this Letter we have proposed a minimal implementation of the PQ 
mechanism in a realistic 
SU(5) model, 
which predicts a narrow axion mass window (cf.~\eq{eq:preferred_window}) which can be directly
tested at future axion DM experiments and indirectly probed by collider and proton decay experiments.
In principle, a precise determination of $m_a$ (via ABRACADABRA and/or CASPEr-Electric) 
would lead to a direct determination of the 
GUT scale,  possibly discriminating among GUT models, and setting a target for proton decay measurements. 
Although we exemplified our predictions in the case of the WGG+$24_F$ model, 
it would be interesting to compare axion properties in other minimal extensions of the WGG model 
which can simultaneously address neutrino masses and gauge coupling unification 
(see e.g.~\cite{Dorsner:2005fq,Dorsner:2005ii}), or in realistic SO(10) models \cite{Ernst:2018bib}.    
   
Finally, the intriguing possibility that the $24_H$ field could also be responsible for inflation 
would make the WGG+$24_F$ model a potential candidate for a minimal and predictive 
GUT-SMASH \cite{Ballesteros:2016euj,Ballesteros:2016xej} variant aiming at a self-contained description of particle physics, 
from the electroweak scale to the Planck scale, and of cosmology, from inflation until today. 
We leave a detailed investigation of this scenario for future studies.

\Sec{Acknowledgments.}
We thank Dmitry Budker, Anne Ernst, Mark Goodsell, Maxim Pospelov, Richard Fibonacci Ruiz, Alex Sushkov, and Lindley Winslow 
for very helpful discussions and communications. C.T. acknowledges support by the Collaborative Research Centre SFB1258 of the Deutsche Forschungsgemeinschaft (DFG).

\bibliographystyle{apsrev4-1.bst}
\bibliography{bibliography_v1}

\end{document}